# Ultrafast Imaging and the "Phase Problem" for Inelastic X-Ray Scattering

By *Peter Abbamonte, Gerard C. L. Wong, David G. Cahill, James P. Reed, Robert H. Coridan, Nathan W. Schmidt, Ghee Hwee Lai, Young Il Joe\*, Diego Casa†*

[*]     Peter Abbamonte, Gerard C. L. Wong, David G. Cahill, James P. Reed, Robert H. Coridan, Nathan W. Schmidt, Ghee Hwee Lai, Young Il Joe
        Frederick Seitz Materials Research Laboratory
        University of Illinois
        Urbana, IL, 61801 (USA)
        E-mail: abbamonte@mrl.uiuc.edu

[†]     Diego Casa
        Advanced Photon Source
        Argonne National Laboratory
        Argonne, IL, 60439 (USA)

We describe a new method for imaging ultrafast dynamics in condensed matter using inelastic x-ray scattering (IXS).  We use the concepts of causality and irreversibility to construct a general solution to the inverse scattering problem (or "phase problem") for inelastic x-ray scattering, which enables direct imaging of dynamics of the electron density with resolutions of $\sim$1 attosecond ($10^{-18}$ sec) in time and < 1 Å in space.  This method is not a Fourier transform of IXS data, but a means to impose causality on the data and reconstruct the charge propagator.  The method can also be applied to inelastic electron or neutron scattering.  We give a general outline of phenomena that can and cannot be studied with this technique, and provide an outlook for the future.

# 1. Ultrafast Dynamics, Irreversibility, and Causality

Many of the most pressing scientific issues in condensed matter science hinge on advancing understanding and ultimately the control of complex interactions among electrons, atoms, and molecules. The microscopic mechanisms and coupling strengths of these interactions can sometimes be inferred from measurements of structure, susceptibilities, and steady-state transport properties, but the complexity of the problems often preclude a complete understanding based on these traditional methods alone. Ultra-fast, time-resolved studies of dynamics can now access molecular and even electronic timescales [1,2,3,4,5]. A central goal in studies of ultrafast dynamics is the determination of cause and effect relationships between observed events, from which one can test models and make predictions. This has enabled highly detailed tests of theory and the discovery of new phenomena that are not apparent in experiments conducted at longer time scales.

The concept of cause and effect, i.e. of "causality", is so intuitive that we often assume it is a fundamental law of nature. However, causality is meaningful only for a large ensemble of particles. For a few-body system, the concept of causality is ill-defined. To illustrate why, consider what would happen if we dropped a ceramic dish above a stone floor. The dish strikes the floor and shatters. It seems obvious, in this case, that the dropping event and shattering event have a causal relationship, i.e. the shattering was "caused" by the dropping, not the other way around.

This one-way relationship, however, is unexpected given Newton's laws, which are "time-reversal invariant". That is, Newton's laws are unchanged even if the flow of time is reversed. For example, a ball thrown through the air, following a parabolic trajectory, will appear to satisfy Newton's laws even if it is recorded on video and played backward. Time reversal invariance is general in physics, and also applies to quantum phenomena, the many-body Schrödinger equation also being time-reversal invariant. [6]

So how can the shattering dish exhibit such clear causality?  In a many-body system, processes that are time-reversal invariant in principle can become irreversible in practice, because of the difficulty in meeting the initial conditions of the time-reversed state.  If we, for example, played the shattering of the dish in reverse, we would see a collection of fragments reassemble into a dish, and then rise into the air.  Clearly, such a process could never occur.  The reason is not that the process violates the laws of physics, but that the set of initial velocities of all the fragments is so unlikely to occur that the probability of ever seeing such an event is infinitely small.

In other words, the concept of causality is not fundamental.  It is only meaningful for a large number of particles.  Our perception of cause and effect, and of time "flowing" in one direction, is nothing but the accumulation of randomness in collections of particles.  Such randomness is captured by the quantity "entropy", whose perpetual rise is dictated by the second law of thermodynamics.  There is, as originally observed by Eddington [7], no microscopic "arrow of time" in nature.

In this article we describe an approach to studies of ultrafast phenomena, based on inelastic x-ray scattering (IXS) using synchrotron x-ray sources, that is complementary to pump-probe methods.  We will explain how the concepts of causality and irreversibility may be used to solve the so-called "inverse scattering problem" for IXS, providing a general strategy for imaging dynamics in many-electron systems.  One of the essential advances has been improvement in technology for synchrotron experiments, which provided the improvement in data quality needed for inversion algorithms to work.  A central advantage of this technique is that it can access extraordinarily short time scales – with temporal resolution of ~ 1 attosecond, and simultaneously access ultra-small length scales, typically less than 1 Å, which in combination provide a fuller way to understand phenomena.

This technique of studying ultrafast phenomena using IXS may seem completely different from the more familiar pump-probe approach.  We will, however, point out a surprisingly precise connection between the two approaches that suggests a deeper meaning of the term "dynamics". We will describe the kinds of phenomena that can be studied with this technique, what phenomena cannot, and where it falls in the spectrum of methods for studying ultrafast dynamics.

## 2. The Propagator

The most general quantity describing the dynamics of a system is its "propagator". To begin, suppose a system is described by the Schrödinger wave equation

$$\left\{ \hat{H}(t) - i\hbar \frac{\partial}{\partial t} \right\} \Psi_n(\mathbf{x}_1, \mathbf{x}_2 \ldots \mathbf{x}_N, t) = 0 \tag{1}$$

where $H$ is a Hamiltonian and $\Psi$ is a many-body wave function describing the system. The retarded propagator, $G_R$, is defined as

$$G_R(\omega) = \frac{1}{\hbar\omega - \hat{H}(0) + i\gamma} \tag{2}$$

where $\gamma$ is a small parameter that prevents the inverse from diverging. If written explicitly in terms of the coordinates and time, the propagator has the form

$$G_R(\mathbf{x}_1', \mathbf{x}_1, \mathbf{x}_2', \mathbf{x}_2 \ldots, t) = -\frac{i}{\hbar} \sum_n \Psi_n^*(\mathbf{x}_1', \mathbf{x}_2' \ldots \mathbf{x}_N', 0) \, \Psi_n(\mathbf{x}_1, \mathbf{x}_2 \ldots \mathbf{x}_N, 0) \exp\left(-i\omega_n t\right) \theta(t) \tag{3}$$

where $\theta(t)$ is a step function. The propagator, physically, describes the probability that the system will evolve from one configuration at time $t = 0$ to another at time $t$.

Unlike the Hamiltonian, $\hat{H}$, the propagator $G_R$ contains information about causality. The analytic properties of $G_R$ are chosen to impose a cause and effect relationship between the excitations of the system, in effect phasing them properly in time. $G_R$ is therefore an explicit measure of the dynamics of the system.

Various conventions for causality may be chosen. For the case of the retarded propagator, $G_R$, the dynamics take place only for $t > 0$, and entropy always increases. One could, alternatively, choose an advanced propagator, $G_A$, which is nonzero only for $t < 0$ and entropy always decreases (i.e., $G_A$ is time time-reversed version of $G_R$). Time-ordered and other conventions may also be used [8]. The convention chosen is arbitrary – similar to the use of retarded versus advanced potentials in electrodynamics – and is a matter of choosing a sign convention for $\gamma$ in eq. 2.

$G_R$ describes everything that can be known about the dynamics of a system. Unfortunately, it is usually impossible to determine completely. A traditional approach, then, is to seek propagators not for the entire system, but specific physical observables, such as the electron density, local magnetic moment, etc., which can be related to $G_R$ through physical models.

## 3. Inelastic X-Ray Scattering (IXS) and its Inverse Problem

Propagators can be determined from scattering experiments. The measured intensity in scattering is proportional to the imaginary part of the propagator for some physical observable [9]. Scattering experiments are, in principle, a direct way to probe the dynamical properties of many-particle systems [9]. IXS, for example, probes the dynamical properties of the total electron density. Inelastic electron scattering, often referred to as electron energy loss spectroscopy (EELS), probes the total charge density (including both electrons and nuclei). Inelastic neutron scattering detects the nuclei and the local magnetic moment.

An IXS experiment is done by directing a collimated, monochromatic beam of x-rays at a specimen, and measuring the intensity of x-rays scattered at different angles and energies. If the incident photon energy is kept far from all of the core absorption levels, resonant processes will be suppressed and the scattered intensity will be proportional to the dynamic structure factor [9,10]

$$S(\mathbf{k}, \omega) = \sum_{m,n} \left| < m \,|\, \hat{n}(\mathbf{k}) \,|\, n > \right|^2 P_n \, \delta(\omega - \omega_m + \omega_n)$$

(3)

Where |n> is an eigenstate of the electronic system, $\omega$ is the energy transferred to the sample by the photon, $\mathbf{k}$ is the momentum transfer, and $P_n = exp[-E_n/k_BT]$ is the Boltzmann factor. $\hat{n}(\mathbf{k})$ is the Fourier transform of $\hat{n}(\mathbf{x},0)$, the operator for the total electron density at $t$=0. The delta function imposes energy conservation.

The dynamic structure factor is the Fourier transform of the Van Hove correlation function for the electron density

$$C(\mathbf{x},t) = \int dx' dt' < \hat{n}(\mathbf{x},t)\hat{n}(\mathbf{x}+\mathbf{x}',t+t') >$$

(4)

where $<\ldots>$ denotes a quantum mechanical, thermal average [10]. $C$ describes the degree to which the electron density at the origin at time $t$=0 is correlated with that at position $\mathbf{x}$ at some later time $t$ (even due to quantum fluctuations at zero temperature). This correlation function, which can be obtained by Fourier transforming the data, contains some information about dynamics, in the sense that its value is determined by the properties of excited states of the system. However, $C$ has no causal properties, so its relation to dynamics is indirect.

Fortunately, $S$ is also related to the propagator for the electron density, by the quantum mechanical version of the Fluctuation-Dissipation theorem,

$$S(\mathbf{k},\omega) = \frac{1}{\pi}\frac{1}{1-e^{\hbar\omega/kT}}\,\text{Im}\left[\chi(\mathbf{k},\omega)\right]$$

(5)

where the retarded propagator $\chi$ is defined as

$$\chi(x,t) = -\frac{i}{\hbar}\left\langle\left[\hat{n}(x,t),\hat{n}(0,0)\right]\right\rangle\theta(t)$$

.

(6)

Here $\hat{n}(\mathbf{x},t)$ is an operator for the total electron density, [,] is a commutator, and $\theta(t)$ is a step function. Physically, $\chi$ describes the probability that a disturbance in the electron density at the origin at time $t$=0 will propagate to position $\mathbf{x}$ at some later time $t$.[11]

Unlike $C(\mathbf{x},t)$, $\chi(\mathbf{x},t)$ is a direct measure of electron dynamics. As a true propagator, $\chi$ is causal, meaning that it properly phases excitations in time. $\chi$ adheres to the second law of thermodynamics, i.e. it always exhibits rising entropy.

Unfortunately, reconstructing $\chi$ from $S$ is not as simple as performing a Fourier transform. The experiment provides only the imaginary part, $Im[\chi]$. We therefore have only a subset of the total data needed. To view the dynamics explicitly, we must solve the so-called "inverse scattering problem" for IXS.

Inverse problems, many of which are colloquially referred to as "phase problems", are innate to all scattering techniques. They are encountered in subjects as diverse as ultrasonic imaging [12], geophysical surveying [13], radar [14], structural biology [15], coastal evolution [16], and in pure mathematics in the study of coupled differential equations [17]. In all cases, the solution involves using preexisting knowledge of the specimen to set constraints that, if combined with the scattering data, yield the missing information. To study dynamics explicitly, we must find such a constraint for the density propagator, $\chi$.

## 4. Causality as a Solution to the Inverse Scattering Problem for IXS

To find a solution to the inverse problem it helps to have a deeper understanding of what information was lost by the experiment. As discussed earlier, propagators (unlike Hamiltonians) exhibit causality. The convention for causality is arbitrary; one may work with a retarded, advanced, time-ordered, or other propagator. The space-time behavior of these propagators is different, e.g. $\chi_R(\mathbf{x},t)$ exhibits rising entropy, and $\chi_A(\mathbf{x},t)$ exhibits falling entropy. However, all these conventions have a common trait: If transformed into $(\mathbf{k},\omega)$ space, they all have the same imaginary part. In reciprocal space, the various causal conventions differ only in their real part.

In other words, the information that is lost in an IXS experiment, which provides only $Im[\chi(\mathbf{k},\omega)]$, is the causality itself. Causality is not definable microscopically, and an IXS experiment provides only microscopic information. To solve the inverse problem, we need a recipe for re-integrating causality with the data from the experiment.

In ref. 18 we presented such a recipe for causally 'ordering' information from an IXS experiment. It involves four steps. The first is to symmetrize the data, to extract the quantity $Im[\chi(\mathbf{k},\omega)]$. This can be done by dividing by the Bose factor in eq. (5), or more cleanly from

$$Im[\chi(\mathbf{k},\omega)] = -\pi\left[S(\mathbf{k},\omega) - S(\mathbf{k},-\omega)\right] \qquad (7)$$

which does not require knowledge of the temperature. Second, one must analytically continue the data, for example with linear interpolation, which is necessary to preserve causality in the final result. Third, one performs a sine transform of the imaginary part, i.e.

$$\chi(\mathbf{k},t) = 2\int d\omega\, Im[\chi(\mathbf{k},\omega)]\cdot\sin(\omega t) \qquad (8)$$

The fourth and final step is a standard spatial Fourier transform from $\chi(\mathbf{k},t)$ to $\chi(\mathbf{x},t)$. This procedure imposes retarded causality onto the data, resulting in a retarded propagator, $\chi(\mathbf{x},t)$, which is nonzero only at $t > 0$ and exhibits increasing entropy.

We emphasize that our procedure is not a simple Fourier transformation of IXS data. Fourier transformation would yield the correlation function, $C(\mathbf{x},t)$, which has no causal properties. Our method is a practical and general solution to the phase problem for IXS, and a way to directly study dynamics of many-particle systems by explicitly determining the charge propagator.

## 5. Experiment

An IXS experiment must be carried out at a synchrotron. Apart from this constraint, the experiment is simple. A typical layout for a moderate energy resolution ($\Delta\omega \sim$ 50-100 meV) experiment is shown in Fig. 1. A synchrotron is an electron storage ring comprising of a lattice of dipole, quadrupole, hexapole and in some cases octapole magnets, separated by straight sections. In these sections reside insertion devices, which are the source of x-rays. The most commonly used insertion device is an undulator, which is a series of dipole magnets whose spacing and field strength are tuned to create constructive interference at specified wavelengths (typically 1 Å) [19]. Its output is a well-collimated, broadband beam of partially coherent hard x-rays, with a bandwidth of ~ 50 eV.

To make this beam suitable for scattering experiments, it is passed through a monochromator comprising two, nondispersively mounted, Bragg reflecting crystals (Fig. 1). The first bounce makes the beam monochromatic and the second redirects it along the beam path. At third generation synchrotrons, where the power density is high, monochromatization is usually done in two steps, a premonochromator that is cooled to absorb the heat load, followed by a secondary monochromator that narrows the bandwidth. This monochromatic beam passes through a series of slits and beam monitors, and strikes the specimen of interest, which is mounted at the center of a goniometer.

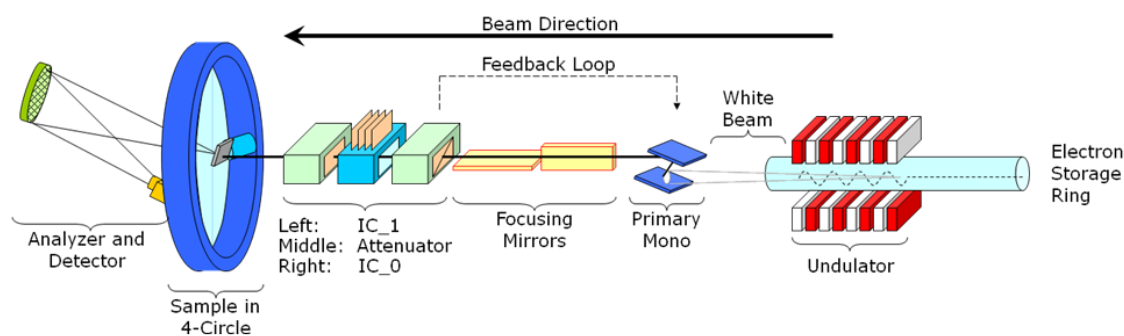

**Figure 1** Schematic of a moderate energy resolution inelastic x-ray scattering setup, such as that at Sector 9 at the Advanced Photon Source.

The scattered x-rays must be energy-analyzed. This is accomplished with an "analyzer", which is mounted on an arm that rotates around the sample position. The analyzer is a mosaic of ~$10^4$ individual mm-sized crystals (typically Si or Ge) mounted on a spherically curved substrate [20,21,22]. The analyzer works in backscattering geometry, reflecting the scattered x-rays onto a photon counting detector next to the sample. The energy transfer $\omega$ is then varied by rotating the angle of the monochromator, and the momentum transfer **k** is varied by rotating the arm and sample goniometer angles. With such a setup one can routinely achieve momentum resolutions of less than 0.1 Å$^{-1}$ and energy resolution less than 100 meV, over a sufficiently broad range of energy and momentum to completely parameterize the cross section in eq. 5. Higher resolutions in energy can be obtained by using dispersively mounted, asymmetric reflections, or high order reflections in backscattering [22]. Higher momentum resolution can be obtained by masking the analyzer or lengthening the arm.

## 6. What Can be Learned from IXS Imaging?

What kind of phenomena can be studied with inverted IXS? An obvious strength is its unprecedented time resolution. As demonstrated in refs. [18,28], IXS imaging can routinely be done with resolutions better than 20 as. By pushing the technology to higher energy, sub-attosecond, i.e., zeptosecond, time-resolution should be achievable. The method therefore allows study of phenomena that are currently beyond the temporal reach of pump-probe techniques.

Another strength of the technique is its simultaneous access to spatial and temporal phenomena at molecular length scales *and* molecular time scales. This allows one to make ultrafast movies of a broad range of physical systems.

Not all phenomena are equally visible with IXS. Recall that its end result is the density propagator, $\chi(\mathbf{x},t) = -i \left\langle \left[ \hat{n}(\mathbf{x},t), \hat{n}(0,0) \right] \right\rangle \theta(t) / \hbar$; $\chi$ reveals excitations that modulate the electron density [18]. IXS is therefore particularly sensitive to longitudinal, collective modes, e.g. plasmons, charge-transfer excitons, and longitudinal phonons. Our original study with this technique involved imaging the valence plasmon in liquid water [18].

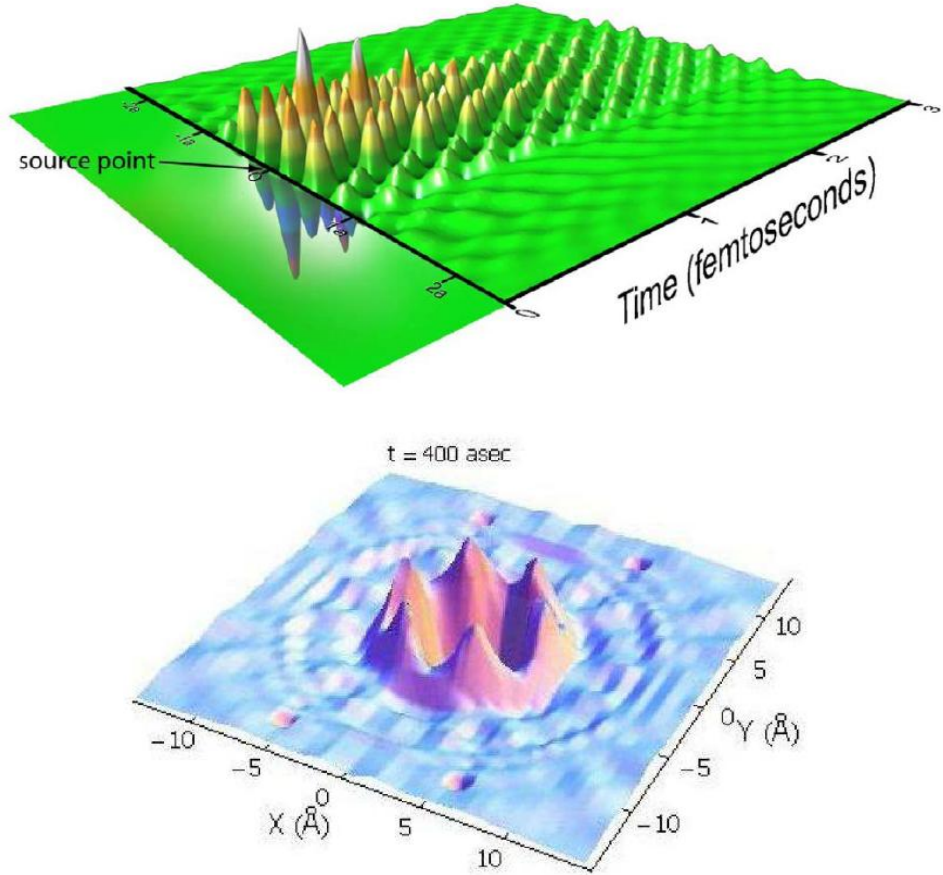

**Figure 2** Examples of charge propagators, $\chi(\mathbf{x},t)$, reconstructed with our solution to the IXS inverse scattering problem. The propagator, physically, corresponds to the disturbance arising from a point source, and the location of this source is visible in these images. (top) $\chi(\mathbf{x},t)$ for LiF, showing time-evolution of the internal structure of the exciton, reproduced from ref. [28]. (bottom) $\chi(\mathbf{x},t)$ for a single crystal of graphite at time $t$=400 as. Visible are plasmons that contribute to the background dielectric constant.

A more recent application, done at Sectors 9 and 15 at the Advanced Photon Source (APS), was our imaging of the internal structure of the exciton in LiF. An exciton is a two-body bound state between a conduction electron and valence hole in an insulator or semiconductor [23]. The first materials in which excitons were observed were the alkali halides, of which LiF is an example [23,28]. The excitons in these systems are of the "charge transfer" type, meaning they involve physical motion of charge from the alkali to the halide atom. This excitation therefore modulates the density and can be observed with IXS. By examining the time-evolution of its internal structure (Fig. 2), we were able to demonstrate that the exciton in LiF is best described by a Frenkel model [28].

Currently we are using inverted IXS to image collective excitations in graphite, using facilities at APS Sector 9 (Fig. 2).  Graphite is a quasi-two-dimensional material with an unusual Fermi surface consisting of small, nested pockets [24].  As a semimetal, graphite has a  low carrier density, but nonetheless fails to exhibit strong correlation effects, i.e. is quite efficient at screening Coulomb interactions.  To understand why, we have been using our images of collective excitations (Fig. 2) to observe the origin of local screening, and explain the surprising ability of graphite to screen charge [25,26,27].

## 7. Analogy with Pump-Probe Techniques

Inverted IXS may seem completely unrelated to the more established pump-probe approach to studying ultrafast dynamics.  The similarities, however, are deeper than they may seem.

To begin, one might think of the propagator itself as a type of pump-probe "experiment".  $\chi(\mathbf{x},t)$ is, by definition, the density disturbance generated by a point source. The source might be thought of as a "pump", the ensuing dynamics being in a sense "probed" by the propagator.

What is the time resolution in this pseudo-pump-probe experiment?   The resolution can be defined in terms of the Nyquist theorem, which gives a time resolution $\Delta t_N = \pi/\Omega_{max}$, where $\Omega_{max}$ is the range of transferred energies measured in the experiment [28].  In ref. 28 we referred to this quantity as a "Nyquist resolution".  The scan range $\Omega_{max}$ therefore, in a manner of speaking, plays the role of the pulse width.

Continuing the analogy, the energy resolution of the IXS experiment, $\Delta\omega$, plays the role of a repetition rate.  To see why, note that the frequency integral eq. (8), at large $t$,  must be done with steps that are much finer than $\Delta\omega$. This necessitates an analytic continuation to preserve causality and prevent leakage of density into negative times [18,28,29]. Continuation should be done with linear interpolation, which creates slope discontinuities in the spectra. These discontinuities, when transformed into time, cause the dynamics to repeat with a period $T = 2\pi/\Delta\omega$, rather like a periodic pump-probe experiment. This repetition sets limits on the duration over which excitations can be examined; phenomena that take place on a time scale longer than $T$ cannot be observed. Slower phenomena can be studied by narrowing the bandwidth $\Delta\omega$, but this lowers the signal and decreases the statistical quality of the data, similar to lowering the repetition rate in a pump-probe experiment.

The broader point is that, ultimately, the most natural way to study dynamics is to perturb a system and examine how it returns to equilibrium. This basic idea is used in both our method and in pump-probe experiments, and the two can be considered different implementations of the same basic conceptual approach.

## 8. Sources and Superposition

The propagator $\chi(\mathbf{x},t)$ represents, physically, the response due to a delta function source at the origin. Because any function can be expressed as a superposition of delta functions, $\chi(\mathbf{x},t)$ permits one to reconstruct – within the assumption of linear response – the disturbance arising from an extended source. The density $n_{ind}(\mathbf{x},t)$ induced by a general time-dependent external charge density $n_{ext}(\mathbf{x},t)$ is given in reciprocal space by

$$n_{ind}(\mathbf{k},\omega) = \frac{4\pi e^2}{k^2} \chi(\mathbf{k},\omega) \, n_{ext}(\mathbf{k},\omega) \tag{9}$$

This expression provides a simple way to model extended, time-dependent sources, properly accounting for coherent interference among excitations.

As an example, we recently used ultra-high resolution IXS data to study hydration structure and dynamics [29]. A molecular understanding of water is central to a broad range of physical phenomena. The short-range structure and dynamics of water has been the subject of intensive study with x-ray absorption spectroscopy [30,31], scattering techniques [32], ultrafast spectroscopy [33,34], such as IR "pump-probe" and vibrational-echo correlation spectroscopy experiments [35,36,37], and molecular dynamics simulations [38]. Much of what we know about hydration comes from this body of experimental and theoretical/computational work on solvation dynamics [39].

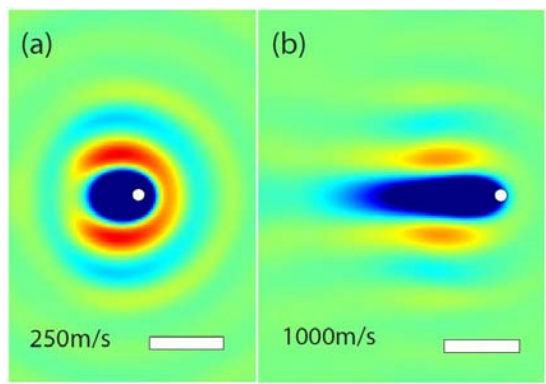

**Figure 3** The hydration structure around a negative charge moving with a steady state linear velocity, calculated from experimental data (see text) (a) v=250m/s, (b) v=1000m/s. Regions of oxygen-density enhancement (indicated by higher electron densities) are red and region of oxygen depletion are blue. The scale bar represents 3Å.

By using a high resolution data set measured at several 3rd generation synchrotron facilities, including 3-ID at the Advanced Photon Source and ID-28 at the ESRF, we were able to reconstruct the charge propagator for liquid water at 26 femtosecond time resolution and 0.44 Å spatial resolution [29]. This time scale is suitable for observing the structural dynamics of water. In fact, IXS has been employed at these facilities to measure spectra reflecting collective dynamics in water [40,41], and the hydration and ionic environments surrounding different biopolymers [42,43]. Using the procedure outlined above, we took these studies a step further and reconstructed the full propagator. The water dynamics we observed is consistent with classical molecular dynamics (MD) simulations of diffusional relaxation, and to femtosecond spectroscopic measurements of the O-O dynamics [29].

To gain insight into hydration, we used this propagator, combined with eq. 9, to produce Fig. 3. This figure shows the 'steady state' dynamical hydration structure around an idealized point negative charge moving with a constant velocity. In the cross-section views (Fig. 3), regions of oxygen-density enhancement around a moving negative point charge are red and region of oxygen depletion are blue. Even for velocities significantly less than the speed of sound in water $v_s$ (≈1600m/s, a typical thermal velocity), the hydration structure deviates from the idealized spherically symmetric "hydration shell" usually assumed, and forms instead a hydration "bowl", with a depletion of oxygen density in front of and especially behind the moving charge distribution. As the velocity increases to 1000 m/s, which is of the same order as $v_s$, the "hydration wake" is formed, and the hydration "bowl" transforms into a hydration "sleeve" (Fig. 3). These results suggest that a stationary molecule and a moving molecule may participate in chemical reactions differently in water.

## 9. What IXS Imaging Cannot Do

Like all experimental techniques, IXS has a range of applicability. Like any other x-ray technique, IXS is more sensitive to phenomena with significant density contrast (see Section 6). Phenomena like spin waves, Wannier excitons, transverse phonons, interband transitions in semiconductors, etc., are difficult to study with IXS as a direct result. Moreover, x-rays are an inherently "weak" probe, meaning that the system responds linearly to the driving field. IXS cannot, therefore, probe the nonlinear optical properties of materials. This makes IXS, for the moment, a passive imaging method. High field physics, second harmonic generation, three- and four- wave mixing, parametric up- or down-conversion are at least for now still relegated to the imagination for x-ray techniques.

The restriction of linear response is more complex when working with extended sources. Returning to the hydration example (see Section 8), linear response dictates that the solvent relaxes through the same modes that govern fluctuations of the interaction at equilibrium. Linear response is believed to hold for most solute-solvent systems, in particular at length scales smaller than those that nucleate a change in phase in the surrounding solvent [44,45,46]. However, it has recently been demonstrated that linear response can fail for cases involving chemical reactions and photoexcitation [47], in which the excitation changes the solute size, affecting the steric state as well as the energetics of hydration, rather than just the latter. Linear response can also break down for phenomena that are so rapid that they break the assumption of continuous solute-solvent interaction.

Limitations from linear response can be overcome in a number of ways. One way is to use a fully self-consistent approach instead of Eq. 9. For example, in the case of an asymmetric solute size, the initial and final charge distribution as well as the initial and final size and shape of the excluded molecular volume may be included in the source itself, the solution being achieved by iteration to convergence. This strategy would mitigate against errors due to breakdown from changes in steric interactions due to changes in molecular shape. Experimentation with such approaches has only just begun.

Finally, IXS imaging – as currently implemented in our algorithms – provides only a spatially averaged response. In a system that lacks translational symmetry, the propagator is actually a function of two momenta, not just one, i.e. $\chi = \chi(\mathbf{k}_1, \mathbf{k}_2, \omega)$. Current IXS methods measure only the diagonal parts of this response, $\chi(\mathbf{k}, \omega) = \chi(\mathbf{k}, -\mathbf{k}, \omega)$. The resulting images are therefore averages over all source locations [25]. For bulk materials, which have some translational symmetry, this is not a serious limitation. However finite systems with no translational symmetry, such as a single molecule, cannot be imaged. This limitation may be overcome with standing wave methods that allow measurement of the so-called "off diagonal" elements of the propagator [18,25].

## 10. Outlook

Progress in improving IXS imaging is accelerating, and many new developments are underway. Standing wave techniques [25] may solve the translational averaging problem (see Section 9), allowing the dynamical study of systems without translational symmetry. Such techniques may open the door to studies of attosecond dynamics of finite objects like patterned devices or biological systems.

We are currently generalizing this IXS technique so that it can treat excluded volumes of finite sized objects, which will allow the reconstruction of liquid dynamics around molecules or in arbitrary geometries. This is achieved via improvement of algorithms for reconstructing the appropriate propagator for a specific geometry from the bulk propagator [29]. These methods will allow us to reconstruct water dynamics in confined geometries, or near surfaces with hydrophobic and hydrophilic domains. Future applications of ultrafast IXS imaging may include biofuels, solar fuels, and ionic liquids, as well as a complete mapping of the Greens function for water, including phases such as supercritical water or supercooled water.

In the long run, we anticipate the lines to blur between laser-based and synchrotron-based approaches to ultrafast dynamics. IXS, currently done at energies ~ 10 keV, traces its origins to nuclear and high energy physics, which over the years have evolved to lower energy to address length and time scales relevant to molecular, biological, and condensed matter phenomena. In parallel, pump-probe methods are being pushed to higher energy and broader bandwidths to reach time scales relevant to fundamental issues of electron dynamics [2,3,4,5]. It seems likely that these two communities will meet at an energy ~ 1 keV, where soft x-ray spectroscopy is particularly useful, resulting in an unprecedented convergence of ultrafast science techniques.

**Acknowledgements**

We acknowledge Thomas Gog, Michael Krisch, and Alfred Baron for important scientific and technical input. This work was supported by the Office of Basic Energy Sciences, U.S. Department of Energy, DE-FG02-07ER46459, through the Frederick Seitz Materials Research Laboratory. Use of the Advanced Photon Source was supported by DOE Contract DE-AC02-O6CH11357. GW is supported by NSF Water CAMPWS and the RPI-UIUC NSEC (DMR-0117792, DMR-0642573).